\documentclass[a4paper,11pt]{article}
\usepackage{jheppub} 
\usepackage{lineno}

\usepackage{xr-hyper}

\makeatletter
\newcommand*{\addFileDependency}[1]{
  \typeout{(#1)}
  \@addtofilelist{#1}
  \IfFileExists{#1}{}{\typeout{No file #1.}}
}
\makeatother

\usepackage{amsmath,amssymb,slashed,braket}
\usepackage{graphicx}
\usepackage{epstopdf}
\usepackage{float,appendix}
\usepackage{esint}

\usepackage[normalem]{ulem}
\usepackage{color}

\definecolor{Orange}{cmyk}{0,0.61,0.87,0}
\definecolor{JungleGreen}{cmyk}{0.99,0,0.52,0}
\definecolor{OliveGreen}{cmyk}{0.64,0,0.95,0.40}
\definecolor{Brown}{cmyk}{0,0.81,1,0.60}
\definecolor{RoyalBlue}{cmyk}{0.71,0.53,0,0.12}
\definecolor{Gray}{cmyk}{0,0,0,0.40}
\definecolor{LightPink}{cmyk}{0.0,0.25,0,0}
\definecolor{LLightPink}{cmyk}{0.0,0.10,0,0}
\definecolor{LightBlue}{cmyk}{0.25,0,0,0}
\definecolor{LightGray}{cmyk}{0,0,0,0.2}

\usepackage{xcolor}
\definecolor{gesfpurple}{rgb}{0.47,0.19,0.42}

\definecolor{gesflanse}{rgb}{0.00,0.50,0.50}

\definecolor{gesfblue}{rgb}{0.08,0.42,0.76}

\definecolor{gesfred}{rgb}{1,0,0}

\definecolor{gesfwhite}{rgb}{1,1,1}

\definecolor{gesfblack}{rgb}{0,0,0}

\newcommand{\geqn}[1]{Eq.\,\hypersetup{linkcolor=blue}(\ref{#1})\hypersetup{linkcolor=blue}}

\graphicspath{{figs/}}

\title{\boldmath Sommerfeld Enhancement from Background Force and the Galactic Center GeV Excess}

\author[*]{Yu Cheng}
\author[*]{and Shuailiang Ge}
\affiliation{Department of Physics, Korea Advanced Institute of Science and Technology (KAIST),\\Daejeon 34141, South Korea}
\note[*]{Corresponding authors.}

\emailAdd{chengyu@kaist.ac.kr}
\emailAdd{shuailiangge@kaist.ac.kr}

\abstract{
We study the impact of background-induced forces on dark matter (DM) annihilation and their implications for indirect detection. In the presence of a finite number density of background particles, loop-level interactions can generate an effective force that is significantly enhanced relative to the vacuum case. We construct a two-component DM model in which the dominant component is a fermionic particle $\chi$ and the subdominant component is an ultralight pseudoscalar particle $\phi$. The annihilation of $\chi$ proceeds through the p-wave channel and produces gamma-ray emission. The finite density of $\phi$ particles induces a background-enhanced force between $\chi$ particles, leading to a sizable Sommerfeld enhancement of the annihilation. We show that a viable region of parameter space in this model can account for the gamma-ray excess observed in the Galactic Center using Fermi-LAT data. The background-induced force substantially amplifies the Sommerfeld enhancement and thus enlarges the parameter space capable of explaining the excess, highlighting the importance of background effects in astrophysical environments.
}

\begin{document}
\maketitle
\flushbottom

\section{Introduction}

More than 80\% of the matter in our Universe today is made of dark matter (DM).
However, the microscopic origin of DM remains unknown, suggesting the existence of new physics beyond the Standard Model (SM) of particle physics.
One major strategy to probe the microscopic nature of DM is indirect detection, which searches for signals from DM annihilation or decay into SM particles.

It has been well recognized that the DM annihilation cross section can be significantly modified in the presence of long-range interactions between DM particles, 
thereby altering the indirect detection signals.
This phenomenon, known as the Sommerfeld enhancement~\cite{Hisano:2002fk,Hisano:2003ec,Hisano:2004ds,Arkani-Hamed:2008hhe,Lattanzi:2008qa,Cirelli:2008pk,Pospelov:2008jd,Fox:2008kb,Pieri:2009zi,Bovy:2009zs,Feng:2009hw,Iengo:2009ni,Cassel:2009wt,Slatyer:2009vg,Yuan:2009bb,Cholis:2010px,Feng:2010zp,Zavala:2009mi,Hryczuk:2011tq,Liu:2013vha,McDonald:2012nc,Zhang:2013qza,Blum:2016nrz, Coy:2022cpt}, arises from the modification of the non-relativistic incoming wavefunction under the long-range attractive potential.
In the presence of such a potential, 
typically of Yukawa or Coulomb type,
the incoming wave is distorted relative to the plane-wave solution in the low-velocity limit,
leading to a substantial enhancement of the short-distance DM annihilation cross section.

The attractive DM self-interaction is usually attributed to tree-level mediator exchange, and the corresponding Sommerfeld enhancement effects on DM annihilation have been extensively studied.
Recently, it has been proposed that analogous enhancement effects can also arise from loop-level quantum forces~\cite{Coy:2022cpt,Ferrante:2025lbs}, mediated by the exchange of two particles.
In particular, if the mediator also serves as a background field
with a large number density, 
the loop-level quantum force can be significantly amplified~\cite{Ferrante:2025lbs}.

This background enhancement originates from the breakdown of the zero-temperature field theory description of the mediator when the number density of background particles becomes sufficiently large.
In calculating the loop-induced force, 
the propagator of the mediator must then be modified to include the on-shell contributions from the background particles.
For example, consider a real scalar (pseudoscalar) mediator $\phi$, 
the expectation value of the normal-ordered product of creation and annihilation operators becomes
\begin{equation}
    \langle n| a_{\boldsymbol{p}} a^\dagger_{\boldsymbol{k}} | n \rangle 
    =
    (2 \pi)^3 \delta^3(\boldsymbol{p} - \boldsymbol{k}) 
    \left[ n(\boldsymbol{k}) +1 \right].
\end{equation}
$n(\boldsymbol{k})$ is the phase space distribution of the DM background. 
This modifies the scalar (pseudoscalar) propagator to be
\begin{equation}
D(k) = 
\frac{i}{k^2 - m_\phi^2 + i \epsilon} + 2 \pi n(\boldsymbol{k})
\delta(k^2 - m_\phi^2),
\label{eq:thermalOP}
\end{equation}
where $m_\phi$ is the mass of the mediator.
The first term is the Feynman propagator in the vacuum, and the second term is the Feynman propagator from a finite-number-density background.
Consequently, a large number density of background particles contributes directly to the loop-level force calculation,
leading to a strongly enhanced effective interaction.
This background effect on the force manifests in two key aspects: first, the force strength increases with the number density of background particles; 
second, the distance dependence of the force is modified. For example, in the case of an axion-mediated spin-independent interaction, the potential changes its scaling behavior from $1/r^3$ to $1/r$~\cite{Cheng:2025fak, Grossman:2025cov}, where $r$ is the distance between two test objects.

The background-induced force was first investigated in the context of neutrino-mediated interactions in the presence of a neutrino background~\cite{Horowitz:1993kw,Ferrer:1998ju,Ferrer:1999ad,Ghosh:2022nzo,Arvanitaki:2022oby,Arvanitaki:2023fij,Ghosh:2024qai,Ittisamai:2025oxf}.
More recently, this framework has been extended to dark forces by exchanging light dark sector particles~\cite{Barbosa:2024pkl,VanTilburg:2024xib}, and specifically to the case of the axion-mediated spin-independent forces between fermions~\cite{Cheng:2025fak, Grossman:2025cov}. Other types of DM background enhancement effects have also been explored~\cite{Fukuda:2021drn,Evans:2023uxh,Arza:2023wou,Evans:2024dty,Zhou:2025wax,Day:2023mkb,Li:2024bbe,Du:2024tin,Yin:2023jjj,Alonso-Alvarez:2019ssa,Gan:2025nlu,Gan:2025icr}.

In this work, we investigate the Sommerfeld enhancement of DM annihilation amplified by background-induced forces and its implications for indirect detection. 
To realize this mechanism, we construct a two-component DM model featuring background-enhanced self-interactions.
An interesting feature of this framework is that the strength of the effective self-interaction is proportional to the number density of background dark matter. 
Consequently, in regions near the Galactic Center (GC) with larger DM densities, the Sommerfeld enhancement becomes more pronounced, leading to an amplified annihilation signal. 
We further show that this mechanism can naturally explain the observed GC gamma-ray excess~\cite{Goodenough:2009gk,Hooper:2010mq,Hooper:2011ti,Abazajian:2012pn,Calore:2014xka,Zhou:2014lva,Fermi-LAT:2015sau, Choquette:2016xsw,Ding:2021zzg,Kong:2025ccv}.

The paper is organized as follows. In Section~\ref{section:two_component_DM}, we introduce a two-component DM model. It is made of a fermionic particle $\chi$ which is the dominant part of DM and an ultralight pseudoscalar particle $\phi$ which is the subdominant part. In Section~\ref{eq:DM_bkg_force}, we derive the effective force between $\chi$ particles induced by the finite-density background of $\phi$ particles. In Section~\ref{section:Bkg_Sommerfeld_effect}, we show how this background-induced force amplifies the Sommerfeld enhancement in $\chi\bar{\chi}$ annihilation. In Section~\ref{section:comparison_with_data}, we compare the gamma-ray spectrum produced by $\chi\bar{\chi}$ annihilation with the observed gamma-ray excess in the GC and show that a viable parameter space of this model can account for the excess. This highlights the phenomenological importance of background-induced forces. Finally, we summarize our results in Section~\ref{section:conclusion}.

\section{Two-component DM model}\label{section:two_component_DM}
We consider a two-component DM model consisting of a Dirac 
fermion $\chi$ and a pseudoscalar $\phi$. The DM abundance is assumed to be dominated by
$\chi$, whose mass lies at the GeV scale, while $\phi$ constitutes a subdominant component with a a sub-eV mass. We define $R$ as the fraction of the relic density of $\chi$ relative to the total DM density $\Omega_{DM}$,
\begin{equation}
    R \equiv \frac{\Omega_\chi}{\Omega_{DM}} =\frac{\Omega_\chi}{\Omega_\chi + \Omega_\phi}
\end{equation}
where $\Omega_{\chi,\phi}$ denotes the relic abundance of $\chi ,\phi$.
We further assume that the light DM component $\phi$ couples to $\chi$ and can therefore mediate a long-range force between $\chi$ particles.
In addition, we introduce a pseudoscalar mediator 
$\eta$ that couples both to 
$\chi$ and Standard Model (SM) fermions. The interaction Lagrangian, together with the corresponding mass terms for each particle, is given by
\begin{equation}
\begin{aligned}
    \mathcal{L} \supset - m_\chi \bar \chi \chi - 
    \frac{1}{2} m_\phi^2 \phi^2 - \frac{1}{2} m_\eta^2 \eta^2 
     + i g_{\phi} \bar \chi \gamma_5 \chi \phi + i g_{\eta} \bar \chi \gamma_5 \chi \eta
     + i g_{f} \bar f \gamma_5 f \eta
\end{aligned}
\end{equation}
where $m_{\chi}$ and $m_\phi$ are the masses of the dominant and subdominant DM components $\chi$ and $\phi$, and $m_\eta$ is the mass of the pseudoscalar mediator $\eta$. 
$g_\phi$ denotes the coupling between $\chi$ and $\phi$, and $g_\eta$ denotes the coupling between $\chi$ and $\eta$. In addition, $g_f$ is the coupling between the mediator $\eta$ and SM fermions $f$.

The dominant DM component $\chi$ is in equilibrium with the SM thermal bath after reheating through its interactions with mediator $\eta$.
Subsequently, $\chi$ gradually freezes out through the p-wave annihilation $\chi \bar \chi \rightarrow \eta \eta$. The evolution of its number density is governed by the Boltzmann equation,
\begin{equation}
\dot{n}_\chi+3 H n_\chi
=
- \left\langle\sigma v_{rel}\right\rangle_{\eta \eta}
\left[
  n_\chi^2
- \left( n^{\rm eq}_\chi \right)^2
\right],
\label{Boltzmann}
\end{equation}
where $n_\chi$ and $n^{eq}_\chi$ are the number density of $\chi$ particle and its equilibrium value. $v_{rel}$ is the relative velocity between $\chi$ and $\bar{\chi}$ particles. The thermal average annihilation cross section for $\chi \bar \chi \rightarrow \eta \eta$ process,  $\left\langle\sigma v_{rel}\right\rangle_{\eta \eta}$, is 
\begin{equation}
\begin{aligned}
(\sigma v_{rel})_{\eta \eta} &=
    \frac{\pi \alpha^2_\eta}{24 m_\chi^2} 
    \sqrt{1 - \frac{m_\eta^2}{m_\chi^2}} v_{rel}^2
    \equiv (\sigma v_{rel})_0 v^2_{rel},
    \\
    \langle \sigma v_{rel} \rangle_{\eta \eta} &=
    \frac{\pi \alpha_{\eta}^2}{4 m_\chi^2} 
    \sqrt{1 - \frac{m_\eta^2}{m_\chi^2}} \frac{1}{x}
    \equiv \langle \sigma  v_{rel} \rangle_0 x^{-1}
    \label{eq:AnnCrosS}
\end{aligned} 
\end{equation}
where $\alpha_\eta \equiv g_\eta^2/4 \pi$ and $x \equiv m_\chi/T$. To obtain the second formula, we have used the definition $\langle \sigma v_{rel} \rangle_{\eta \eta} = x^{3 / 2}/(2 \sqrt{\pi}) \int_0^{\infty}  (\sigma v_{rel})_{\eta \eta}  e^{-x v_{rel}^2/4} v_{rel}^2 d v_{rel}$ and have taken the Taylor expansion in terms of $v_{rel}^2$ in the non-relativistic limit. Following the standard procedure outlined in \cite{Kolb:1990vq}, the relic abundance of $\chi$ can be calculated as
\begin{equation}
 \Omega_\chi h^2
 =
 \frac{\rho_\chi}{\rho_c} h^2= 1.07 \times 10^9 \frac{(n+1) x_f^{(n+1)}}{\left(g_{*s} / g_{*}^{1 / 2}\right) M_{\mathrm{Pl}}
 \langle \sigma  v_{rel} \rangle_0} \mathrm{GeV}^{-1}.   
\end{equation}
with $n = 1$ for a p-wave process and $g_{*s}$ and $g_*$ are the effective degrees of freedom for entropy and energy, respectively. $x_f$ denotes the freeze-out temperature. 
In our scenario, $\chi$ constitutes a fraction $R$ of the total DM, $\Omega_{\chi} = R \, \Omega_{\rm{DM}}$. To match with the current observation of the DM abundance $\Omega_{\rm{DM}} h^2 \simeq 0.12$, the corresponding coupling is  
\begin{equation}
\label{eq:alpha_eta}
\alpha_\eta \sim 0.03 R^{-1/2} \left(\frac{m_\chi}{100\,\rm{GeV}}\right).
\end{equation}

The subdominant dark matter component $\phi$ is assumed to be very light, thus its relic abundance can only be generated non-thermally. Possible production mechanisms include freeze-in, misalignment, etc. To ensure that $\phi$ does not thermalize with the SM thermal bath, the annihilation rate of $\chi \bar \chi \rightarrow \phi \phi$ should remain smaller than the Hubble expansion rate 
\begin{equation}
    n_\chi \langle \sigma v_{rel} \rangle_{\phi \phi} < H \propto \frac{T^2}{M_{pl}} \quad \text{with} \quad \langle \sigma v_{rel} \rangle_{\phi \phi} =
    \frac{\pi \alpha_{\phi}^2}{4 m_\chi^2} \frac{1}{x}
\end{equation}
where $M_{pl}$ is the Planck mass and $\alpha_\phi \equiv g^2_\phi / 4 \pi$. This condition can be conservatively satisfied by choosing $\alpha_\phi < 10^{-7} \sim 10^{-8}$, depending on the mass of $\chi$.
With this constraint on $\alpha_\phi$, the interaction between $\chi$ and $\bar{\chi}$ via exchange of a virtual $\phi$ is strongly suppressed and cannot induce significant Sommerfeld enhancement in the $\chi \bar \chi \rightarrow \eta \eta$ annihilation. Compared with the coupling $\alpha_{\eta}$ in \eqref{eq:alpha_eta}, we see that $\alpha_{\phi} \ll \alpha_{\eta}$.

However, the presence of abundant $\phi$ particles in the early Universe could still induce a large background-enhanced force between $\chi$ and $\bar{\chi}$, thereby significantly promoting the Sommerfeld enhancement in the $\chi \bar \chi \rightarrow \eta \eta$ annihilation. This promotion could substantially modify the freeze-out dynamics of $\chi$ and invalidate Eq.~\eqref{eq:AnnCrosS}. To avoid such complications in the early-Universe evolution, we assume that $\phi$ particles are generated sufficiently late, for example after the epoach of the Big Bang Nucleosynthesis (BBN).
As a result, their abundance remains negligible during the freeze-out epoch, so that they do not affect either the freeze-out process of $\chi$ or its relic abundance. Also, the sufficiently low abundance of $\phi$ does not significantly affect the relic abundance of $\chi$ after the freeze-out. 
In addition, the p-wave annihilation is highly sensitive to the velocity of $\chi$. In the early Universe, the velocity of $\chi$ is low after its freeze-out, 
so the channel $\chi \bar \chi \rightarrow \eta \eta$ could be inefficient even in the presence of Sommerfeld enhancement and background enhancement. 
We will come back to this point in Section~\ref{section:Bkg_Sommerfeld_effect}.
Consequently, we do not need to worry about the evolution of those particles in the early Universe.
In contrast, in the present Universe, the number density of $\phi$ particles can become sufficiently large in regions such as the inner part of the Galaxy,
making their background effects on $\chi$ annihilation non-negligible. The velocity of $\chi$ particles also becomes large in those regions. 
The focus of this work is therefore to study how a finite number density of background $\phi$ particles in the current Galaxy promotes the $\chi$ annihilation and the resulting phenomenological consequences.

\section{DM self-interaction from background effect}
\label{eq:DM_bkg_force}
In our model, the background-enhanced self-interaction between DM $\chi$ can be generated at the loop level either via the exchange of a virtual $\eta$ and a DM particle $\phi$, or through the exchange of a $\phi$ pair. Since $\alpha_\phi \ll \alpha_\eta$, which is required by the condition that $\phi$ never gets equilibrium with the SM thermal bath, the force from $\phi$ pair exchange is highly suppressed by a factor of $\alpha_\phi^2$. Therefore, in the remainder of this paper, we focus primarily on the background-enhanced force arising from the exchange a virtual $\eta$ and a DM particle $\phi$.

\begin{figure}[!t]
    \centering
    \includegraphics[width=0.7
    \textwidth]{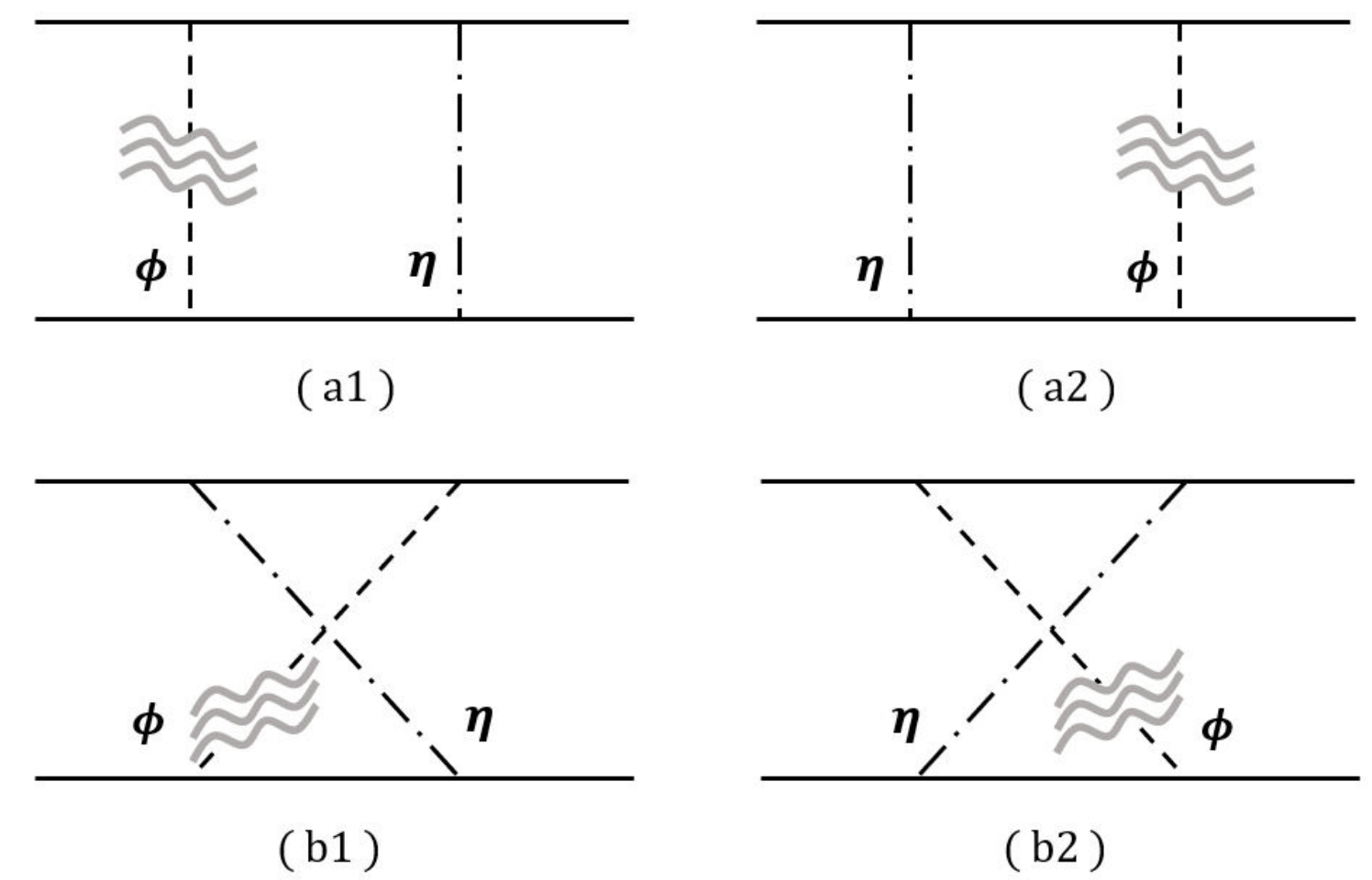}
    \caption{Feynman diagrams which contribute to the self-interacting potential of $\chi$ via the exchange of a virtual mediator $\eta$ and a background DM particle $\phi$. The wavy lines represent the $\phi$ DM background. The background force can be obtained by substituting the $\phi$ propagators with the background modified propagator in Eq.~\eqref{eq:thermalOP}. 
    }
    \label{fig:FeynmanD}
\end{figure}

We now briefly summarize the procedure for computing such a background force between 
$\chi$. 
The relevant Feynman diagrams are shown in Fig.~\ref{fig:FeynmanD}.
This force can be viewed as arising from one fermion $\chi$ scattering off the static potential generated by another fermion $\chi$. In the non-relativistic limit, the potential in the  momentum space, $\tilde{V}(\boldsymbol{q})$, is related to the scattering amplitude via $i \mathcal{M} = - i \tilde{V} (\boldsymbol{q}) 4 m^2_\chi \delta_{s s^\prime} \delta_{r r^\prime}$, where $\boldsymbol{q}$ denotes the momentum transfer between the two $\chi$ particles.
There are totally four Feynman diagrams contributing to the background force in this model, as shown in Fig. \ref{fig:FeynmanD}. In calculating the matrix of these diagrams, one should use the modified propagator for the DM $\phi$, as given in \geqn{eq:thermalOP}. In \geqn{eq:thermalOP}, the first term is the vacuum propagator, and the second term is the propagator from the finite-density of background DM particles $\phi$ which are on shell. Since we are interested in the background contribution to the force, we retain only the crossing terms involving one virtual $\eta$ and one on-shell $\phi$ from the DM background, which is far larger than the contribution from the pure vacuum case with both $\eta$ and $\phi$ being virtual. In addition, the case of both mediator lines being on-shell $\phi$ does not contribute, because the on-shell conditions for the two lines cannot be satisfied simultaneously.

The resultant expression for $\tilde{V} (\boldsymbol{q})$ from the four different diagrams are given by
\begin{eqnarray}
    \tilde{V}_{a1}(\boldsymbol{q}) + \tilde{V}_{a2}(\boldsymbol{q}) = 
    \tilde{V}_{b1}(\boldsymbol{q}) + \tilde{V}_{b2}(\boldsymbol{q}) 
    = \frac{g_\phi^2 g_\eta^2}{4 m_\chi^2}
    F(\boldsymbol{q})
\end{eqnarray}
where 
\begin{equation}
\begin{aligned}
F(\boldsymbol{q})
&\equiv
\int \frac{d^3 k}{(2 \pi)^3}
    \frac{ n(\boldsymbol{k}) }{E_k} \mathcal{A}(\boldsymbol{k},\boldsymbol{q},\Delta),
\\
     \mathcal{A}(\boldsymbol{k},\boldsymbol{q},\Delta) &\equiv
\frac{1}{-\boldsymbol{q}^2 + 2 \boldsymbol{k} \cdot \boldsymbol{q} - \Delta^2} 
+ 
\frac{1}{-\boldsymbol{q}^2 - 2 \boldsymbol{k}\cdot \boldsymbol{q} - \Delta^2}.
\end{aligned}
\end{equation}
with $\Delta \equiv m_\eta^2 - m^2_\phi$.
The corresponding potential in position space is then given by the Fourier transform,
\begin{equation}
    V( \boldsymbol{r})
    =
     \int \frac{d^3 \boldsymbol{q}}{(2 \pi)^3} e^{i \boldsymbol{q} \cdot r} \tilde{V}(\boldsymbol{q}).
     \label{eq:PotentialFourier}
\end{equation}
Assuming an isotropic distribution function $n(k)$ for DM $\phi$, the angular part of momentum integral can be carried out. The resulting  background force between $\chi$ particles can then be expressed as
\begin{equation}
V(\boldsymbol{r})
=
 - \frac{g^2_\phi g_\eta^2}{ 8 \pi^3 m_\chi^2 r^2}
    \int d k |k|
    \frac{n(k)}{E_k}
    \sin(|k| r) e^{-\sqrt{\Delta^2 - |k|^2} r}.   
\end{equation}
For details of the computations of the matrix elements and the derivation of the coordinate-space potential between DM $\chi$ particles in this model, please refer to the Appendix.
To gain further insight into the behavior of the background force, we consider the case where the DM $\phi$ follows the isotropic Maxwell-Boltzman distribution
\begin{equation}
    n(\boldsymbol{k}) = n_{\phi} \frac{(2 \pi)^{3/2}}{\sigma_k^3} 
    e^{-\frac{\boldsymbol{k}^2}{2 \sigma_k^2}}
\end{equation}
where $\sigma_k$ represents the velocity dispersion of the DM and $n_\phi$ is the number density of the DM $\phi$.
Then the expression for the background potential can be simplified to 
\begin{equation}
\begin{aligned}
V(\boldsymbol{r})
&=
- \frac{g_\phi^2 g_\eta^2}{4 \pi m_\chi^2 r}
\frac{n_{\phi}}{m_{\phi}} e^{-\frac{1}{2} \sigma_k^2 r^2 } 
e^{-\Delta r}\\
&=
- \frac{4 \pi \alpha_\phi \alpha_\eta}{m_\chi^2 r}
 \frac{\rho_{\phi}}{m_{\phi}^2}  e^{-m_\eta r} \equiv - \frac{\alpha_{bkg}}{r} e^{-m_\eta r}.
\label{eq:ExpPotentialMaxwell}
\end{aligned}
\end{equation}
where
\begin{equation}
    \alpha_{bkg} \equiv 
    \frac{4 \pi \alpha_\phi \alpha_\eta \rho_\phi}{m_\chi^2 m_\phi^2}.
\end{equation}
We have used the assumption that $m_\eta \gg m_\phi > \sigma_k \sim k$, such that $\Delta \sim m_\eta$. The exponential suppression is dominated by the factor  $e^{-m_\eta r}$. The quantity $\rho_\phi$ denotes the energy densities of DM $\phi$ and is related to the total DM energy density by $\rho_\phi = (1-R) \rho_{DM}$.
From \geqn{eq:ExpPotentialMaxwell}, we find that the background potential behaves as a Yukawa-type potential with the effective coupling $\alpha_{bkg}$, and the force range $\sim 1/m_\eta$. This short-range behavior arises because the virtual $\eta$ propagator is significantly off-shell, which suppresses the background contribution at large distance. As a result, the potential decays exponentially beyond the length $1/m_\eta$. Nonetheless, the force strength is significantly enhanced by the large occupation
number of $\phi$ particles in the background. As $m_\phi$ decreases, this enhancement becomes even more pronounced, resulting in a stronger force.

\section{Background-force-induced Sommerfeld enhancement}
\label{section:Bkg_Sommerfeld_effect}

The attractive background force between DM $\chi$ particles can induce the Sommerfeld enhancement in the DM annihilation process $\chi \bar \chi \rightarrow \eta \eta$.
The enhancement factor $S$ is then defined as the ratio between the enhanced annihilation and the original tree-level cross section.
This enhancement occurs because the presence of the potential modifies the wave function of the incoming particles near the origin, increasing the probability of annihilation compared to the case with no interaction. For the background-enhanced potential of the form given in \geqn{eq:ExpPotentialMaxwell}, the modified wave function is  determined by solving the following Schrödinger equation
\begin{equation}
- \frac{1}{m_\chi} \nabla^2 \psi_k  - \frac{\alpha_{bkg}}{r} e^{-m_\eta r} \psi_k  = m_\chi v^2 \psi_k,
\end{equation}
with the boundary condition $\psi_k \rightarrow e^{i k z} + f(\theta) e^{ikr}/r$ as $r \rightarrow \infty$. 
Here, $v \equiv v_{rel}/2$ is the DM velocity in the center of mass frame. The general solution can be expanded into partial waves as
$\psi_k = \sum_l A_l P_l(\cos \theta) \chi_l(r)/r$ where $P_l(\cos \theta)$ is the Legendre polynomial and $\chi_l(r)/r$ is the radial wave function for angular momentum $l$. 
Following the notations in \cite{Feng:2010zp,Tulin:2013teo}, we introduce the dimensionless parameters $\epsilon_v \equiv v/\alpha_{bkg}$, $\epsilon_\phi \equiv m_\phi/(\alpha_{bkg} m_\chi)$ and $\tilde{r} \equiv \alpha_{bkg} m_\chi r$. 
In terms of these variables, the radial wave function $\chi_l(\tilde{r})$ can be obtained by solving the differential equation,
\begin{equation}
    \frac{d^2 \chi_l (\tilde{r})}{d \tilde r^2} + \left( \epsilon_v^2 - \frac{l(l+1)}{\tilde{r}^2} + \frac{1}{\tilde{r}}e^{-\epsilon_\phi \tilde{r}} \right) \chi_l (\tilde{r}) = 0,
    \label{eq:DiffEqRadialWave}
\end{equation}
with the boundary conditions $\chi_{l}(\tilde{r})=\left(\epsilon_v \tilde{r}\right)^{l+1}$ when $\tilde{r} \rightarrow 0$ and $\chi_{l}(\tilde{r}) \rightarrow C_{l} \sin \left(\epsilon_v \tilde{r}-\frac{l \pi}{2}+\delta_{l}\right)$ when $\tilde{r} \rightarrow \infty$. $C_l$ is a normalization constant and $\delta_l$ is the phase shift.
The Sommerfeld enhancement factor for a generic $l$ partial wave is given by \cite{Iengo:2009ni,Cassel:2009wt,Ding:2021zzg}
\begin{equation}
    S_l = \left| \frac{\chi_l(0)}{\chi^{(0)}_l (0)}\right|^2 =\left[\frac{(2 l+1)!!}{C_{l}}\right]^2
\end{equation}
where $\chi^{(0)}_l$ denotes the free solution to \geqn{eq:DiffEqRadialWave} in the absence of the potential.

An analytic solution for the Sommerfeld enhancement factor for the Yukawa potential can be obtained by approximating it by the Hulthen potential \cite{Iengo:2009ni}. Under this approximation, the enhancement factors for s-wave ($l=0$) and p-wave ($l=1$) annihilations, $S_s$ and $S_p$, are given by \cite{Iengo:2009ni,Cassel:2009wt,Tulin:2013teo,Ding:2021zzg}
\begin{equation}
\begin{aligned}
     S_s &= \frac{\pi}{\epsilon_v} 
    \frac{\sinh \left(\frac{2 \pi \epsilon_v}{\pi^2 \epsilon_{\phi} / 6}\right)}{\cosh \left(\frac{2 \pi \epsilon_v}{\pi^2 \epsilon_{\phi} / 6}\right)
    -\cos \left(2 \pi \sqrt{\frac{1}{\pi^2 \epsilon_{\phi} / 6}
    -\frac{\epsilon_v^2}{\left(\pi^2 \epsilon_{\phi} / 6\right)^2}}\right)},\\
     S_p 
     &=
    \frac{\left(1-\varepsilon_{\phi} \pi^2 / 6\right)^2
    +4 \varepsilon_v^2}{\left(\varepsilon_{\phi} \pi^2 / 6\right)^2+4 \varepsilon_v^2} S_s.
\end{aligned}
\end{equation}

Now, we consider the Sommerfeld enhancement on the DM annihilation $\chi \bar \chi \rightarrow \eta \eta$ within our Galactic halo. According to \geqn{eq:ExpPotentialMaxwell}, the strength of the background-induced self-interaction between DM $\chi$ particles depends on the energy density of DM component $\phi$, with the relation $\alpha_{bkg} \propto \rho_{\phi}$. Consequently, the Sommerfeld enhancement factor $S$ is also a function of $\rho_\phi$ and thus a function of the distance from the GC.
Both DM components are assumed to follow the NFW profile,
\begin{equation}
 \rho_{DM}(r)
 =
 \rho_s \frac{\left(r / r_s\right)^{-\gamma}}{\left(1+r / r_s\right)^{3-\gamma}}.
\end{equation}
$r$ denotes the distance from the GC and $r_s = 20\,$kpc is a reference scale. The constant $\rho_s = 0.3371\,\rm{GeV}/\rm{cm}^3$ is fixed by requiring that the local DM density at the Sun's position, $r_\odot = 8.5\,$kpc, satisfies $\rho_\phi\left(r_{\odot}\right) = 0.43\,\rm{GeV}/\rm{cm}^3$. For the canonical NFW profile, we have $\gamma =1$, while in the generalized NFW profile, $\gamma$ is treated as a free parameter which typically
lies within the range $\gamma \in (0.9, 1.3)$.
The energy densities of the two DM components are then given by $\rho_\chi = R \, \rho_{DM}$ and $\rho_\phi = (1 - R) \,\rho_{DM}$.

\begin{figure}[!t]
    \centering
    \includegraphics[width=0.8
    \textwidth]{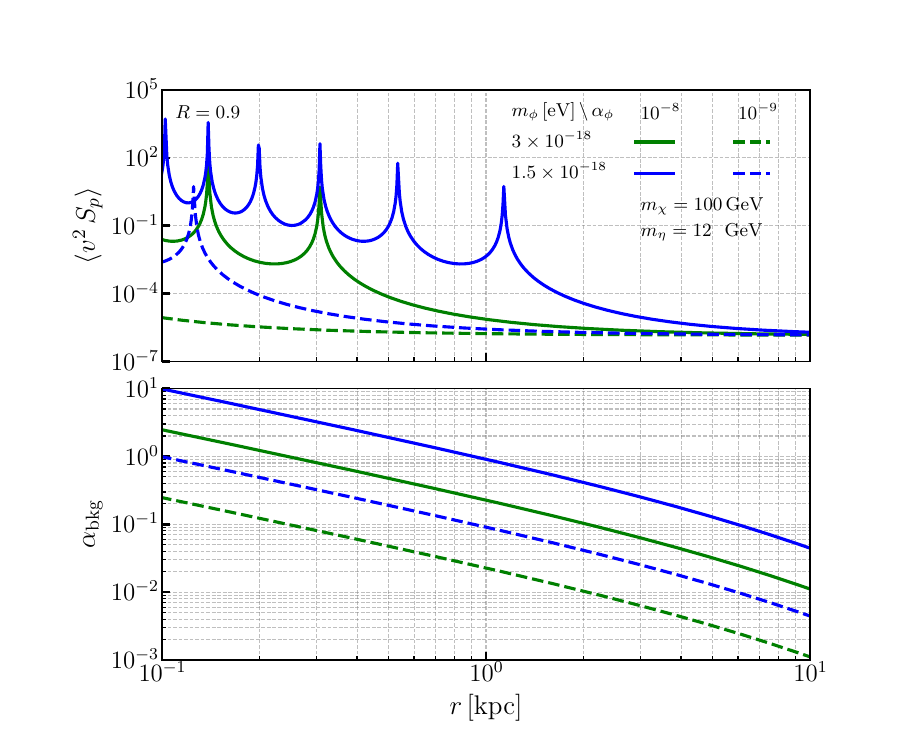}
    \caption{ The effective Sommerfeld enhancement factor $\langle v^2 S_p \rangle$ (upper panel) and the effective coupling $\alpha_{bkg}$ (lower 
    panel) as a function of the distance $r$ from the GC. We choose four benchmark points of the parameters $(m_{\phi},\alpha_{\phi})$, which are $m_\phi = 3 \times 10^{-18}\,$eV and $1.5 \times 10^{-18}\,$eV, and $\alpha_\phi = 10^{-8}$ and $10^{-9}$. In addition, for illustrative purposes, we fix $m_\chi = 100\,$GeV and  
    $m_\eta = 12\,$GeV, and take the fraction of the dominant DM component $\chi$ to be $R = 0.9$.
    }
    \label{fig:SAlphaBkg}
\end{figure}

The $\chi \bar \chi \rightarrow \eta \eta$ annihilation is p-wave dominant, whose cross section is proportional to $v^2$, the square of the DM $\chi$'s velocity. This motivates us to define an effective enhancement factor $\langle v^2 S_p \rangle$. After performing an average over the DM velocity distribution, we obtain
\begin{equation}
\langle v^2 S_p \rangle  =
\frac{1}{v_0^3} \sqrt{\frac{2}{\pi}} \int_0^{\infty} d v_{\mathrm{rel}} v_{\mathrm{rel}}^4 S_p  e^{- \frac{v_{\rm rel}^2}{2 v_0^2} }.
\end{equation}
Here, $v_0 = 105\,\rm{km}/\rm{s}$ is the 1D velocity dispersion. 
Then, the Sommerfeld-enhanced 
annihilation cross-section can be written as 
\begin{equation}
\langle \sigma v_{\rm rel} \rangle_S = (\sigma v_{\rm rel})_0 \langle v^2 S_p \rangle.
\end{equation}
$(\sigma v_{\rm rel})_0$ is
the velocity-independent part of the cross section without Sommerfeld enhancement, as given in \geqn{eq:AnnCrosS}. Fig.~\ref{fig:SAlphaBkg} shows the dependence of both the effective Sommerfeld enhancement factor $\langle v^2 S_p \rangle$ and the 
coupling strength $\alpha_{bkg}$ on the distance $r$ from the GC. We have chosen four benchmark points of the parameters $(m_{\phi}, \alpha_{\phi})$ in making the plot. From the figure, we can see that both $\langle v^2 S_p \rangle$ and $\alpha_{bkg}$ increase as the distance $r$ decreases. This behavior arises from the scaling $\rho_{DM} \propto 1/r$ near the GC. Furthermore, a smaller $m_\phi$, which corresponds to a higher number density, leads to a stronger background-induced self-interaction potential between $\chi$ particles, resulting in a more significant Sommerfeld enhancement of the DM annihilation.

The background-induced Sommerfeld enhancement can also occur during the epoch of recombination and potentially affect the Cosmic Microwave Background (CMB) power spectrum. 
However, the velocity of DM $\chi$ at that epoch is highly red-shifted after freeze-out, which is expected to scale as
\begin{eqnarray}
    v \propto \sqrt{\frac{T_{\rm kd}}{m_\chi}} \frac{T_{\rm CMB}}{T_{\rm kd}} \sim 10^{-8} - 10^{-10}
\end{eqnarray}
where $T_{\rm kd}$ and $T_{\rm CMB}$ are the temperatures at the times of kinetic decoupling and recombination, respectively, which leads to a highly suppressed effective Sommerfeld enhancement factor $\langle v^2 S_p \rangle$.
As a result, the corresponding annihilation rate remains negligible, allowing the model to easily evade current CMB constraints.

In addition, the two-component dark matter model, with one heavy and one light component, may lead to deviations from the standard NFW profile. In particular, the background-induced self-interaction between $\chi$ particles could give rise to a core profile in the inner region of the Galaxy. However, since the background force is both density-dependent and time-oscillating, a precise determination of the resulting density profile would require dedicated N-body simulations that incorporate the dynamics of both components. Such an analysis lies beyond the scope of this work, and we therefore adopt the NFW profile as a simplifying assumption.

\section{DM indirect detection with Fermi-LAT}
\label{section:comparison_with_data}
In the outer region of the galaxy halo, the DM annihilation cross section is highly suppressed due to its p-wave nature, resulting in negligible signals for indirect
detection. However, as shown in Fig.~\ref{fig:SAlphaBkg}, within a few kpc from the GC, the annihilation cross section is substantially enhanced by the background-force-induced Sommerfeld enhancement effect. 
The $\eta$ particles produced from DM $\chi$ annihilation
will further promptly decay into SM fermions and then subsequently decay into gamma rays as an observable signal.

The photon flux generated by the DM $\chi$ annihilation in a given direction can be expressed as 
\begin{equation}
 \Phi\left(E_\gamma, l, b\right)
 =
 \frac{1}{4} \frac{(\sigma v_{rel})_0}{4 \pi m_\chi^2} \frac{d N_\gamma}{d E_\gamma} J_S(l, b),   
\end{equation}
where $l$ and $b$ denote the longitude and latitude in the galactic coordinate, and
$(\sigma v_{rel})_0$ is the velocity-independent part of the annihilation cross section, as defined in \geqn{eq:AnnCrosS}. $d N_\gamma / d E_\gamma$ is the gamma-ray spectrum produced per DM annihilation. In our model, we assume the $\eta$  mediator primarily decay into $b \bar b$ pairs
\footnote{The coupling of pseudoscalar $\eta$ with SM fermions is generally proportional to the fermion mass, such that $\eta$ predominantly decays into the heaviest SM fermion allowed by kinematics. For a benchmark value of $m_\eta = 12\,$GeV, the dominant decay channel is $b \bar b$.}, which then decay into gamma-ray photons. We use the {\tt PPPC4DMID}
package \cite{Cirelli:2010xx} to generate the photon spectrum and further
boost it to the galactic frame.
The quantity $J_S$ is the effective $J$ factor, defined as the integral of the squared DM density along the line of-sight, incorporating the effective Sommerfeld enhancement factor, 
\begin{equation}
\begin{aligned}
    J_S(l, b)
    =
    \int_0^{\infty} R^2
    \langle v^2 S_p \rangle
    \left(\sqrt{s^2-2 r_{\odot} s \cos l \cos b+r_{\odot}^2}\right) \\
    \times \rho_{DM}^2\left(\sqrt{s^2-2 r_{\odot} s \cos l \cos b+r_{\odot}^2}\right) d s.
\end{aligned}
\end{equation}
$s$ parametrizes 
the line-of-sight distance, $r_{\odot} = 8.5\,$kpc denotes the distance from the Sun to the GC. The total gamma ray flux can then be obtained by integrating $\Phi(E_\gamma,l,b)$ over the relevant Galactic coordinates $l,b$ in the observational region.

\begin{figure}[!t]
    \centering
    \includegraphics[width=0.8
    \textwidth]{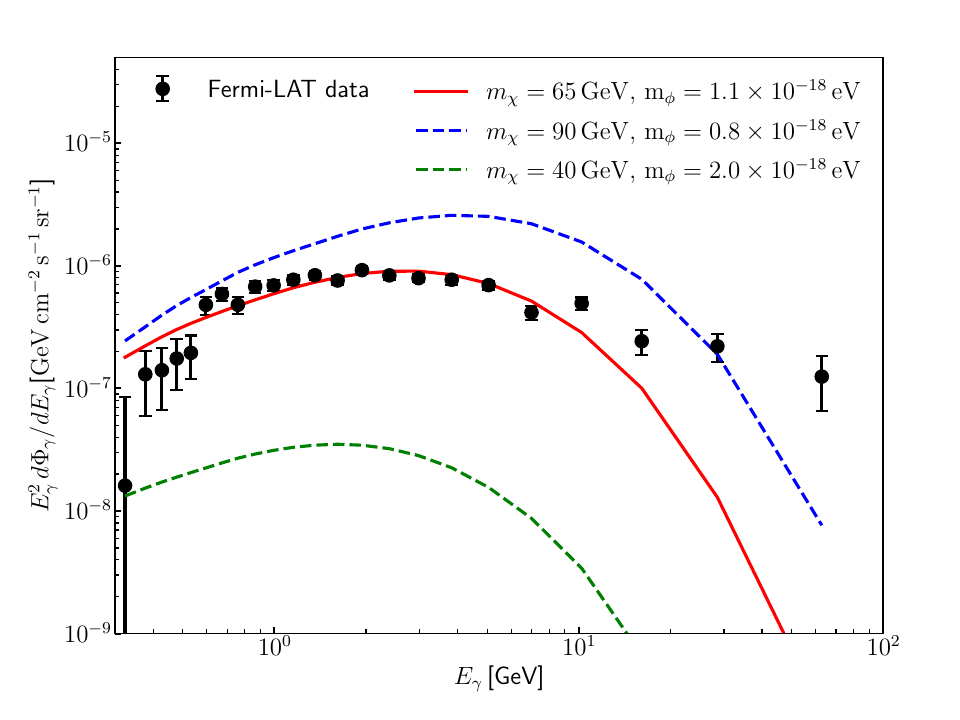}
    \caption{ 
    The Fermi-LAT data (black points with error bar) and the best-fit result with $m_\chi = 65\,$GeV and $m_\phi = 1.1 \times 10^{-18}\,$GeV (red solid line). For comparison, two alternative parameter sets are also displayed: $m_\chi = 90\,$GeV, $m_\phi = 0.8 \times 10^{-18}\,$GeV (blue dashed line) and $m_\chi = 40\,$GeV, $m_\phi = 2 \times 10^{-18}\,$GeV (green dashed line). 
     For illustration, the remaining model parameters are fixed as $\alpha_\phi = 10^{-8}$, $m_\eta = 12\,$GeV and the DM $\chi$ fraction $R = 0.9$.
    }
    \label{fig:FitData}
\end{figure}

We use the result of Galactic Center excess (GCE) data in \cite{Calore:2014xka} and compared it with the gamma-ray flux predicted in our model. We adopt the region of interest of $|l|< 20^{\circ}$ and $2^{\circ}<|b|<20^{\circ}$, which is specified in \cite{Calore:2014xka}, with the model parameters fixed as $\alpha_\phi = 10^{-8}$, $m_\eta = 12\,$GeV, $R = 0.9$. After performing the chi-squared minimization, we find the best-fit point at $m_\chi = 65\,$GeV and $m_\phi = 1.1 \times 10^{-18}\,$GeV, with the corresponding spectrum shown as the red solid line in Fig. \ref{fig:FitData}. 
For comparison, we also show the predicted gamma ray flux for two other parameter choices:  $m_\chi = 90\,$GeV, $m_\phi = 0.8 \times 10^{-18}\,$GeV and $m_\chi = 40\,$GeV, $m_\phi = 2 \times 10^{-18}\,$GeV. 

From Fig.~\ref{fig:FitData}, we see that a certain parameter space of our model can account for the gamma-ray excess in the GC. The novel feature in our model is that we have considered the background-induced effect. The effective Sommerfeld enhancement factor can reach $\mathcal{O}(10^2)$ once the background-induced force is taken into account, as illustrated in Fig.~\ref{fig:SAlphaBkg}. This sizable enhancement substantially enlarges the viable parameter space capable of explaining the gamma-ray excess observed in the GC. This effect is particularly important for the p-wave annihilation channel, which is less efficient for indirect detection due to its intrinsic $v^2$ suppression. In our framework, the key ingredient is the background-enhanced force mediated by the abundant ultralight $\phi$ particles, whose number density increases toward the GC. Since the effective $\alpha_{\rm bkg}$ is proportional to the local energy density of $\phi$, the Sommerfeld enhancement factor becomes strongly position-dependent and grows rapidly in regions of high DM density. This naturally leads to an annihilation rate that is significantly boosted in the inner Galaxy while remaining suppressed in the outer halo.  
Moreover, the ultralight nature of the $\phi$ field implies that the background-induced force is inherently space and time dependent, as the energy density of $\phi$ particles fluctuates
\footnote{
The ultralight $\phi$ field is $\phi(x) = \phi_0 \cos(\omega_{\phi} t - k_{\phi}x)$ where $\omega_{\phi}\simeq m_{\phi}$ since it is non-relativistic with $v_{\phi}\sim 10^{-3}$. Then, the energy density is calculated as $\rho_{\phi} = \frac{1}{2}\dot{\phi}^2 + \frac{1}{2}(\nabla \phi)^2 + \frac{1}{2}m_{\phi}\phi^2  \simeq \frac{1}{2}m_{\phi}^2\phi_0^2 + \Delta \rho_{\phi}$ where the energy density fluctuation is $\Delta \rho_{\phi} = \frac{1}{2}(\nabla \phi)^2 = \frac{1}{2}k_{\phi}^2\phi_0^2\cos^2(\omega_{\phi}t - k_{\phi}x)$.
} at the level of $\Delta \rho_{\phi}/\rho_{\phi} \sim v_{\phi}^2$ in a time scale set by $\sim 1/m_{\phi}$ and in a space scale set by $\sim 1/(v_{\phi}m_{\phi})$.
As a result, the induced Sommerfeld enhancement is also expected to exhibit temporal and spatial modulation. Although the timescale of this oscillation is shorter than current observational integration times for gamma-ray observations, and the space scale of this oscillation is smaller than the spatial resolution of the observations, this feature represents a qualitatively novel prediction of the model. 
In principle, such space and time dependent effects could leave imprints in precision observations or in future analyses sensitive to temporal and spatial variations.

\section{Conclusion}
\label{section:conclusion}

In this work, we have investigated the impact of background-induced forces on DM annihilation and their phenomenological consequences for indirect detection. To realize this mechanism in a concrete and controlled framework, we constructed a two-component DM model consisting of a dominant fermionic DM component $\chi$ and an ultralight pseudoscalar component $\phi$. While $\chi$ undergoes annihilation into a heavy mediator $\eta$ through p-wave process, the presence of a finite-density background of ultralight $\phi$ particles significantly enhances the effective self-interaction between $\chi$ particles. This background-enhanced force therefore induces a sizable Sommerfeld enhancement of the $\chi\bar{\chi}$ annihilation rate.

A feature of our framework is the focus on p-wave annihilation. The p-wave process is typically regarded as less efficient due to its intrinsic $v^2$ suppression. In our scenario, this suppression plays a role in ensuring consistency with the early-Universe constraints. After freeze-out, the velocity of $\chi$ particles is strongly red-shifted, rendering the Sommerfeld-enhanced p-wave annihilation negligible during cosmological epochs such as recombination. Furthermore, we assume that the ultralight $\phi$ particles are generated sufficiently late, so that they do not affect the freeze-out process and the relic abundance of $\chi$ significantly. In contrast, the background-induced Sommerfeld enhancement could become relevant in the late Universe, particularly in galactic environments where the DM density is high and the velocities of $\chi$ particles is relatively large. 

In the present Galaxy, the energy density of $\phi$ particles increases toward the GC, leading to a position-dependent enhancement of the effective coupling governing the long-range force between $\chi$ particles. We have shown that this effect can produce an effective Sommerfeld enhancement factor as large as $\mathcal{O}(10^2)$ in the inner Galaxy. This substantially enlarges the viable parameter space capable of accounting for the observed GeV gamma-ray excess in the GC. Our results demonstrate that incorporating background-induced forces provides a novel and efficient mechanism for explaining DM signals in certain astrophysical environments.

An additional distinctive aspect of our model arises from the ultralight nature of the $\phi$ field. The background energy density of $\phi$ exhibits intrinsic temporal and spatial modulations, which are inherited by the effective Sommerfeld enhancement factor. Although these modulations occur on time and length scales that are currently below the resolution of gamma-ray observations and are therefore averaged out in existing analyses, they represent a qualitatively new prediction of the framework. In principle, such space and time dependent effects could become relevant in future precision studies or in complementary observational probes sensitive to time-resolved or small-scale variations.

In summary, this work highlights the importance of background effects in astrophysical environments and demonstrates that they can play a crucial role in DM phenomenology. The mechanism studied here could be important in exploring detection signatures of DM. More broadly, it motivates further investigations into the interplay between background-induced forces and observable signals in a variety of astrophysical contexts.

\acknowledgments
We would like to thank Yi-Ming Zhong for helpful discussions. The work is supported by the National Research Foundation of Korea (NRF) Grant RS-2023-00211732, by the Samsung Science and Technology Foundation under Project Number SSTF-BA2302-05, by the POSCO Science Fellowship of POSCO TJ Park Foundation, and by the National Research Foundation of Korea (NRF) Grant RS-2024-00405629.

\appendix

\section{Feynman diagram calculation}

\begin{figure}[h]
    \centering
    \includegraphics[width=0.9
    \textwidth]{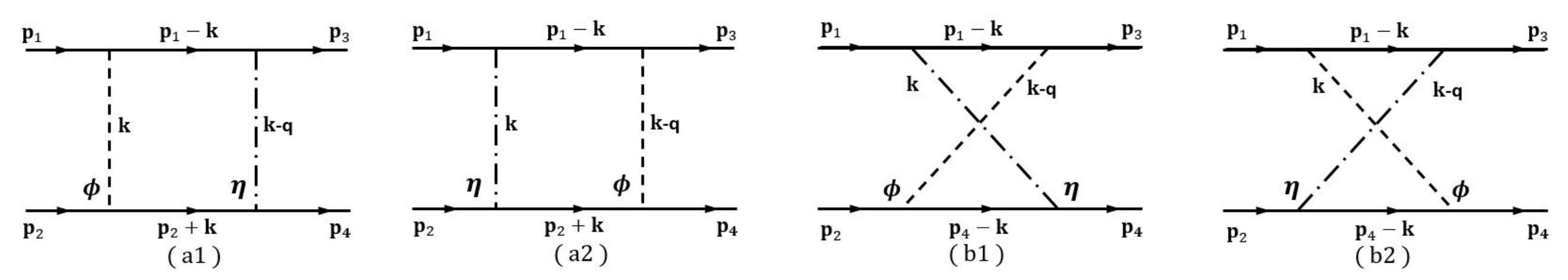}
    \caption{The Feynman diagrams correspond to the background-enhanced force  by exchanging one virtual $\eta$ and one DM $\phi$ in our two-component DM model.}
    \label{fig:Feyndetail}
\end{figure}

In this Appendix, we provide a detailed calculation of the background-enhanced force by exchanging one virtual $\eta$ and one  DM $\phi$ in our two component DM model.
A total of four Feynman diagrams contribute to the potential, as shown in Fig. \ref{fig:Feyndetail}. 
The external momenta of the Feynman diagrams are labeled as $p_1$, $p_3$ corresponding to the initial and final states of DM particles $\chi$, and $p_2$, $p_4$ for anti-particle $\bar \chi$. The arrows on fermion lines represent the momentum directions. The momenta for two internal lines are labeled as $k$ and $k-q$, where $q \equiv p_1 - p_3 \approx (0, \boldsymbol{q}) $ represents the momentum transfer during the scattering process.

As an example, we start with computing the potential $\tilde{V}_{a1} + \tilde{V}_{a2}$; the other two diagrams can be obtained in a similar way. The matrix element for diagrams $a1)$ and $a2)$ can be written as 
\begin{equation}
    \begin{aligned}
&i \left( \mathcal{M}_{a1} + \mathcal{M}_{a2} \right) \\
&= 
g_\phi^2 g_\eta^2
\int \frac{d^4 k }{(2 \pi)^4}
\bar u(p_3) 
 \gamma_5 \frac{i \left( \slashed{p}_1 - \slashed{k} + m_{\chi}
\right)}{(p_1 - k)^2 - m_{\chi}^2} 
     \gamma_5 u(p_1)  \bar v(p_2) 
     \gamma_5 \frac{i \left( -(\slashed{p}_2 + \slashed{k}) + m_{\chi}
    \right)}{(p_2 + k)^2 - m_{\chi}^2}  \gamma_5 v(p_4)\\
&\times \left[
    \left(\frac{i}{k^2 - m_\phi^2} + 2 \pi n(k) \delta(k^2 - m_\phi^2)\right)
    \frac{i}{(k-q)^2 - m_\eta^2} \right. \\
    &\left.\quad +  
    \frac{i}{k^2 - m_\eta^2} \left(\frac{i}{(k-q)^2 - m_\phi^2} + 2 \pi n(k-q) \delta((k-q)^2 - m_\phi^2)\right)\right].
\end{aligned}
\end{equation}
We have kept only the crossing terms which have one $\delta$ function since we are interested in the background contribution, while neglecting the pure vacuum contribution. Using the identities for fermions, $\slashed{p}_{1,3} u(p_{1,3}) = m_\chi u(p_{1,3})$ and $\slashed{p}_{2,4} v(p_{2,4}) = - m_{\chi} v(p_{2,4})$ with
$\bar u (p_{3}) \slashed{q} u (p_{1})  = 0$ and $\bar v (p_{2}) \slashed{q} v (p_{4})  = 0$, we obtain
\begin{equation}
\begin{aligned}
i \left( \mathcal{M}_{a1} + \mathcal{M}_{a2} \right) 
 &=
 - \int \frac{d^4 k }{(2 \pi)^4}
g_\phi^2 g_\eta^2
\left( 2 \pi n(k) \delta(k^2 - m_\phi^2)\right)\\
&\times
\left[
\bar u(p_3) 
\frac{  \slashed{k} }{(p_1 - k)^2 - m_{\chi}^2} 
u(p_1) \bar v(p_2) 
\frac{ \slashed{k} }{(p_2 + k)^2 - m_{\chi}^2} 
v(p_4)  \frac{i}{(k-q)^2 - m_\eta^2}  \right. \\
&\left. +
\bar u(p_3) 
\frac{  \slashed{k}  }{(p_3 - k )^2 - m_{\chi}^2} 
u(p_1) \bar v(p_2) 
\frac{ \slashed{k} }{(p_4 + k )^2 - m_{\chi}^2} 
v(p_4)
\frac{i}{(k+q)^2 - m_\eta^2}
\right].
\end{aligned}
\end{equation}
Due to the on-shell conditions guaranteed by the $\delta$ function in the integral, 
\begin{equation}
    \delta(k^2 - m_\phi^2) =  \frac{1}{2 E_k} 
    \left[ \delta(k^0 - E_k) +  \delta(k^0 + E_k) \right],
    \label{eq:DeltaFRel}
\end{equation}
the denominator in the integral can be simplified to
\begin{equation}
    \begin{aligned}
    (p_{1,3} - k)^2 - m_{\chi}^2 
    &= 
   m_\phi^2 \mp 2 E_k m_{\chi} + 2 \boldsymbol{k}\cdot \boldsymbol{p}_{1,3} 
   \approx
   \mp 2 E_k m_{\chi},\\
    (p_{2,4} + k)^2 - m_{\chi}^2 
   &=
   m_\phi^2 \pm 2 E_k m_{\chi} - 2 \boldsymbol{k}\cdot \boldsymbol{p}_{2,4}
   \approx
    \pm 2 E_k m_{\chi}.
    \end{aligned}
\end{equation}
The sign in front of $E_k$ is determined by satisfying either the first or the second $\delta$ function in \geqn{eq:DeltaFRel}. In the derivation, we take $|\boldsymbol{k}| \ll m_\phi \ll m_{\chi}$, as the high $\boldsymbol{k}$ contribution to the integral is exponentially suppressed by the DM momentum distribution. This allows the spatial component of the vector product $\boldsymbol{k} \cdot \boldsymbol{p}$ to be safely neglected. 
Recalling in the non-relativistic limit, we have $\bar{u}_{s^\prime} 
\left(p_3\right) \gamma^\mu u_s \left(p_1\right)
= 2 m_{\chi} \delta_0^\mu \delta_{s^\prime s}$ and 
$\bar{v}_{r^\prime} \left(p_2\right) \gamma^\mu v_r \left(p_4\right)= - 2 m_{\chi} \delta_0^\mu \delta_{r^\prime r}$. We can integrate out $k^0$ and obtain
\begin{equation}
\begin{aligned}
&i \left( \mathcal{M}_{a1} + \mathcal{M}_{a2} \right) \\
&=
g_\phi^2 g_\eta^2
\int \frac{d^4 k}{(2 \pi)^4}
\frac{ 2 \pi n(k) }{2 E_k}
\left(
\delta(k_0 - E_k) + \delta(k_0 + E_k)
\right)
\\
&\times
\left[
4 m^2_{\chi} \delta_{s^\prime s} \delta_{r^\prime r}
\frac{  k_\mu \delta^{\mu}_0 }{\mp 2 E_k m_{\chi}} 
\frac{  k_\nu \delta^{\nu}_0 }{\pm 2 E_k m_{\chi}}
\left(
\frac{i}{k^2 + q^2 - 2 k q - m_\eta^2} +
\frac{i}{k^2+q^2+2kq - m_\eta^2}\right)
\right]\\
    &=
    -i
    \left(4 m^2_{\chi} \delta_{s^\prime s} \delta_{r^\prime r}\right)
    \frac{g_\phi^2 g_\eta^2}{4 m^2_{\chi}} 
    \int \frac{d^3 k}{(2 \pi)^3}
    \frac{ n(k) }{E_k}
    \left[
    \frac{1}{-\boldsymbol{q}^2 + 2 \boldsymbol{k} \cdot \boldsymbol{q} - \Delta^2} 
    + 
    \frac{1}{- \boldsymbol{q}^2 - 2 \boldsymbol{k}\cdot \boldsymbol{q} - \Delta^2}
    \right]
\end{aligned}
\end{equation}
where $\Delta^2 \equiv m_\eta^2 - m_\phi^2$. Then, using the relation $i \mathcal{M} = - i \tilde{V} (\boldsymbol{q}) 4 m^2_{\chi} \delta_{s s^\prime}\delta_{r r^\prime}$, the potential is determined as
\begin{equation}
    \tilde{V}_{a1}(\boldsymbol{q}) + \tilde{V}_{a2}(\boldsymbol{q}) 
    =
    \frac{g_\phi^2 g_\eta^2}{4 m^2_{\chi}}
    \int \frac{d^3 k}{(2 \pi)^3}
    \frac{ n(\boldsymbol{k}) }{E_k}
    \left[
    \frac{1}{-\boldsymbol{q}^2 + 2 \boldsymbol{k} \cdot \boldsymbol{q} - \Delta^2} 
    + 
    \frac{1}{- \boldsymbol{q}^2 - 2 \boldsymbol{k}\cdot \boldsymbol{q} - \Delta^2}
    \right].
\end{equation}

\section{Coordinate-space potential from Fourier transformation}

Next, we preform Fourier transformation to obtain the background-induced potential in coordinate space. Here, we take the general form of the momentum-space potential,
\begin{eqnarray}
 \tilde{V}_i(q) = A_i \int \frac{d^3 k}{(2 \pi)^3} \frac{n(k)}{E_k}
 \left[
\frac{1}{-\boldsymbol{q}^2 + 2 \boldsymbol{k} \cdot \boldsymbol{q} 
- \Delta^2 - i \epsilon} 
+ 
\frac{1}{- \boldsymbol{q}^2 - 2 \boldsymbol{k}\cdot \boldsymbol{q}
- \Delta^2 + i \epsilon}
\right]
\end{eqnarray}
where $A_i$ is an arbitrary coefficient and the $i \epsilon$ prescription corresponds to the retarded propagator.
Applying the Fourier transformation in \geqn{eq:PotentialFourier} and shifting the momentum integration variables $\boldsymbol{q} \rightarrow \boldsymbol{q} + \boldsymbol{k}$ for the first term and $\boldsymbol{q} \rightarrow \boldsymbol{q} - \boldsymbol{k}$ for the second term inside the bracket, we obtain
\begin{equation}
    \begin{aligned}
    V_i(r) 
    &=
    A_i
    \int \frac{d^3 k}{(2 \pi)^3} 
    \frac{n(k)}{E_k}
     \int \frac{d^3 q}{ (2 \pi)^3}
    \left[ 
    \frac{ e^{ i \boldsymbol{(q + k)} \cdot \boldsymbol{r}} }{- \boldsymbol{q}^2 -  (\Delta^2 -\boldsymbol{k}^2) - i \epsilon}
    + \frac{ e^{ i \boldsymbol{(q-k)} \cdot \boldsymbol{r}} }{- \boldsymbol{q}^2 - ( \Delta^2 - \boldsymbol{k}^2 ) + i \epsilon} \right] \\
    &=
    - A_i
    \int \frac{d^3 k}{(2 \pi)^3} 
    \frac{n(k)}{E_k}
     \frac{1}{ 4 \pi |\boldsymbol{r}|}  e^{ -\sqrt{ \Delta^2 - \boldsymbol{k}^2} |\boldsymbol{r}|}
    \left(
    e^{ i \boldsymbol{k} \cdot \boldsymbol{r}}
    +
    e^{ -i \boldsymbol{k} \cdot \boldsymbol{r}}
    \right)\\
    &=
    - \frac{A_i}{2 \pi r}
    \int \frac{d^3 k}{(2 \pi)^3} 
    \frac{n(k)}{E_k}
    \text{Re} \left[ e^{i \boldsymbol{k} \cdot \boldsymbol{r}}
    e^{ -\sqrt{ \Delta^2 - \boldsymbol{k}^2} |\boldsymbol{r}|}\right]
    \end{aligned}
\end{equation}
where we assume $\Delta^2 \gg \boldsymbol{k^2}$.
If the phase space distribution function $n(k)$ is isotropic, the angular integration simplifies the expression further to
\begin{eqnarray}
    V_i (r) 
    =
    - \frac{A_i}{ 4 \pi^3 r^2}
    \int d k |k|
    \frac{n(k)}{E_k}
    \sin(|k| r) e^{-\sqrt{\Delta^2 - |k|^2} r}.
    \label{eq:ExpPotentialISO}
\end{eqnarray}
%



\bibliographystyle{JHEP}
\bibliography{ref.bib}

@article{Cirelli:2010xx,
    author = "Cirelli, Marco and Corcella, Gennaro and Hektor, Andi and Hutsi, Gert and Kadastik, Mario and Panci, Paolo and Raidal, Martti and Sala, Filippo and Strumia, Alessandro",
    title = "{PPPC 4 DM ID: A Poor Particle Physicist Cookbook for Dark Matter Indirect Detection}",
    eprint = "1012.4515",
    archivePrefix = "arXiv",
    primaryClass = "hep-ph",
    reportNumber = "CERN-PH-TH-2010-057, SACLAY-T10-025, IFUP-TH-2010-44",
    doi = "10.1088/1475-7516/2012/10/E01",
    journal = "JCAP",
    volume = "03",
    pages = "051",
    year = "2011",
    note = "[Erratum: JCAP 10, E01 (2012)]"
}

@article{Calore:2014xka,
    author = "Calore, Francesca and Cholis, Ilias and Weniger, Christoph",
    title = "{Background Model Systematics for the Fermi GeV Excess}",
    eprint = "1409.0042",
    archivePrefix = "arXiv",
    primaryClass = "astro-ph.CO",
    reportNumber = "FERMILAB-PUB-14-289-A",
    doi = "10.1088/1475-7516/2015/03/038",
    journal = "JCAP",
    volume = "03",
    pages = "038",
    year = "2015"
}

@book{Kolb:1990vq,
    author = "Kolb, Edward W. and Turner, Michael S.",
    title = "{The Early Universe}",
    reportNumber = "FERMILAB-BOOK-1990-01",
    doi = "10.1201/9780429492860",
    isbn = "978-0-201-62674-2",
    volume = "69",
    year = "1990"
}

@article{Ferrer:1998ju,
    author = "Ferrer, F. and Grifols, J. A. and Nowakowski, M.",
    title = "{Long range forces induced by neutrinos at finite temperature}",
    eprint = "hep-ph/9806438",
    archivePrefix = "arXiv",
    reportNumber = "UAB-FT-448",
    doi = "10.1016/S0370-2693(98)01489-0",
    journal = "Phys. Lett. B",
    volume = "446",
    pages = "111--116",
    year = "1999"
}

@article{Grossman:2025cov,
    author = "Grossman, Yuval and Yu, Bingrong and Zhou, Siyu",
    title = "{Axion forces in axion backgrounds}",
    eprint = "2504.00104",
    archivePrefix = "arXiv",
    primaryClass = "hep-ph",
    month = "3",
    year = "2025"
}

@article{VanTilburg:2024xib,
    author = "Van Tilburg, Ken",
    title = "{Wake forces in a background of quadratically coupled mediators}",
    eprint = "2401.08745",
    archivePrefix = "arXiv",
    primaryClass = "hep-ph",
    doi = "10.1103/PhysRevD.109.096036",
    journal = "Phys. Rev. D",
    volume = "109",
    number = "9",
    pages = "096036",
    year = "2024"
}

@article{Evans:2023uxh,
    author = "Evans, Jason L.",
    title = "{Effect of Ultralight Dark Matter on g-2 of the Electron}",
    eprint = "2302.08746",
    archivePrefix = "arXiv",
    primaryClass = "hep-ph",
    doi = "10.1103/PhysRevLett.132.091801",
    journal = "Phys. Rev. Lett.",
    volume = "132",
    number = "9",
    pages = "091801",
    year = "2024"
}

@article{Arza:2023wou,
    author = "Arza, Ariel and Evans, Jason",
    title = "{Electron $g-2$ corrections from axion dark matter}",
    eprint = "2308.05375",
    archivePrefix = "arXiv",
    primaryClass = "hep-ph",
    month = "8",
    year = "2023"
}

@article{Evans:2024dty,
    author = "Evans, Jason L. and Lyu, Jun-Yuan",
    title = "{Effective two-loop background contributions to $g_e-2$}",
    eprint = "2410.10715",
    archivePrefix = "arXiv",
    primaryClass = "hep-ph",
    month = "10",
    year = "2024"
}

@article{Ghosh:2022nzo,
    author = "Ghosh, Mitrajyoti and Grossman, Yuval and Tangarife, Walter and Xu, Xun-Jie and Yu, Bingrong",
    title = "{Neutrino forces in neutrino backgrounds}",
    eprint = "2209.07082",
    archivePrefix = "arXiv",
    primaryClass = "hep-ph",
    doi = "10.1007/JHEP02(2023)092",
    journal = "JHEP",
    volume = "02",
    pages = "092",
    year = "2023"
}

@article{Ghosh:2024qai,
    author = "Ghosh, Mitrajyoti and Grossman, Yuval and Tangarife, Walter and Xu, Xun-Jie and Yu, Bingrong",
    title = "{The neutrino force in neutrino backgrounds: Spin dependence and parity-violating effects}",
    eprint = "2405.16801",
    archivePrefix = "arXiv",
    primaryClass = "hep-ph",
    doi = "10.1007/JHEP07(2024)107",
    journal = "JHEP",
    volume = "07",
    pages = "107",
    year = "2024"
}

@article{Ferrer:1999ad,
    author = "Ferrer, F. and Grifols, J. A. and Nowakowski, M.",
    title = "{Long range neutrino forces in the cosmic relic neutrino background}",
    eprint = "hep-ph/9906463",
    archivePrefix = "arXiv",
    reportNumber = "UAB-FT-468, IFT-P-053-99",
    doi = "10.1103/PhysRevD.61.057304",
    journal = "Phys. Rev. D",
    volume = "61",
    pages = "057304",
    year = "2000"
}

@article{Horowitz:1993kw,
    author = "Horowitz, C. J. and Pantaleone, James T.",
    title = "{Long range forces from the cosmological neutrinos background}",
    eprint = "hep-ph/9306222",
    archivePrefix = "arXiv",
    reportNumber = "IUHET-249, IUNTC93-14",
    doi = "10.1016/0370-2693(93)90800-W",
    journal = "Phys. Lett. B",
    volume = "319",
    pages = "186--190",
    year = "1993"
}

@article{Arvanitaki:2022oby,
    author = "Arvanitaki, Asimina and Dimopoulos, Savas",
    title = "{Cosmic neutrino background on the surface of the Earth}",
    eprint = "2212.00036",
    archivePrefix = "arXiv",
    primaryClass = "hep-ph",
    doi = "10.1103/PhysRevD.108.043517",
    journal = "Phys. Rev. D",
    volume = "108",
    number = "4",
    pages = "043517",
    year = "2023"
}

@article{Arvanitaki:2023fij,
    author = "Arvanitaki, Asimina and Dimopoulos, Savas",
    title = "{A Diffraction Grating for the Cosmic Neutrino Background and Dark Matter}",
    eprint = "2303.04814",
    archivePrefix = "arXiv",
    primaryClass = "hep-ph",
    month = "3",
    year = "2023"
}

@article{Barbosa:2024pkl,
    author = "Barbosa, Sergio and Fichet, Sylvain",
    title = "{Background-induced forces from dark relics}",
    eprint = "2403.13894",
    archivePrefix = "arXiv",
    primaryClass = "hep-ph",
    doi = "10.1007/JHEP01(2025)021",
    journal = "JHEP",
    volume = "01",
    pages = "021",
    year = "2025"
}

@article{Zhou:2025wax,
    author = "Zhou, Kevin",
    title = "{Ponderomotive Effects of Ultralight Dark Matter}",
    eprint = "2502.01725",
    archivePrefix = "arXiv",
    primaryClass = "hep-ph",
    month = "2",
    year = "2025"
}

@article{Fukuda:2021drn,
    author = "Fukuda, Hajime and Shirai, Satoshi",
    title = "{Detection of QCD axion dark matter by coherent scattering}",
    eprint = "2112.13536",
    archivePrefix = "arXiv",
    primaryClass = "hep-ph",
    reportNumber = "IPMU21-0089",
    doi = "10.1103/PhysRevD.105.095030",
    journal = "Phys. Rev. D",
    volume = "105",
    number = "9",
    pages = "095030",
    year = "2022"
}

@article{Day:2023mkb,
    author = "Day, Hannah and Liu, Da and Luty, Markus A. and Zhao, Yue",
    title = "{Blowing in the dark matter wind}",
    eprint = "2312.13345",
    archivePrefix = "arXiv",
    primaryClass = "hep-ph",
    doi = "10.1007/JHEP07(2024)136",
    journal = "JHEP",
    volume = "07",
    pages = "136",
    year = "2024"
}

@article{Li:2024bbe,
    author = "Li, Shao-Ping and Xie, Ke-Pan",
    title = "{Photon proliferation from N-body dark matter annihilation}",
    eprint = "2412.15749",
    archivePrefix = "arXiv",
    primaryClass = "hep-ph",
    month = "12",
    year = "2024"
}

@article{Du:2024tin,
    author = "Du, Mingxuan and Liu, Jia and Wang, Xiao-Ping and Wu, Tianhao",
    title = "{Enhanced monochromatic photon emission from millicharged co-interacting dark matter}",
    eprint = "2403.07528",
    archivePrefix = "arXiv",
    primaryClass = "hep-ph",
    doi = "10.1007/JHEP10(2024)026",
    journal = "JHEP",
    volume = "10",
    pages = "026",
    year = "2024"
}

@article{Yin:2023jjj,
    author = "Yin, Wen",
    title = "{Thermal production of cold \textquotedblleft{}hot dark matter\textquotedblright{} around eV}",
    eprint = "2301.08735",
    archivePrefix = "arXiv",
    primaryClass = "hep-ph",
    reportNumber = "TU-1177",
    doi = "10.1007/JHEP05(2023)180",
    journal = "JHEP",
    volume = "05",
    pages = "180",
    year = "2023"
}

@article{Alonso-Alvarez:2019ssa,
    author = "Alonso-\'Alvarez, Gonzalo and Gupta, Rick S. and Jaeckel, Joerg and Spannowsky, Michael",
    title = "{On the Wondrous Stability of ALP Dark Matter}",
    eprint = "1911.07885",
    archivePrefix = "arXiv",
    primaryClass = "hep-ph",
    reportNumber = "IPPP/19/84",
    doi = "10.1088/1475-7516/2020/03/052",
    journal = "JCAP",
    volume = "03",
    pages = "052",
    year = "2020"
}

@article{Feng:2010zp,
    author = "Feng, Jonathan L. and Kaplinghat, Manoj and Yu, Hai-Bo",
    title = "{Sommerfeld Enhancements for Thermal Relic Dark Matter}",
    eprint = "1005.4678",
    archivePrefix = "arXiv",
    primaryClass = "hep-ph",
    reportNumber = "UCI-TR-2010-06",
    doi = "10.1103/PhysRevD.82.083525",
    journal = "Phys. Rev. D",
    volume = "82",
    pages = "083525",
    year = "2010"
}

@article{Tulin:2013teo,
    author = "Tulin, Sean and Yu, Hai-Bo and Zurek, Kathryn M.",
    title = "{Beyond Collisionless Dark Matter: Particle Physics Dynamics for Dark Matter Halo Structure}",
    eprint = "1302.3898",
    archivePrefix = "arXiv",
    primaryClass = "hep-ph",
    doi = "10.1103/PhysRevD.87.115007",
    journal = "Phys. Rev. D",
    volume = "87",
    number = "11",
    pages = "115007",
    year = "2013"
}

@article{Ding:2021zzg,
    author = "Ding, Yu-Chen and Ku, Yu-Lin and Wei, Chun-Cheng and Zhou, Yu-Feng",
    title = "{Consistent explanation for the cosmic-ray positron excess in p-wave Sommerfeld-enhanced dark matter annihilation}",
    eprint = "2104.14881",
    archivePrefix = "arXiv",
    primaryClass = "hep-ph",
    doi = "10.1088/1475-7516/2021/09/005",
    journal = "JCAP",
    volume = "09",
    pages = "005",
    year = "2021"
}

@article{Iengo:2009ni,
    author = "Iengo, Roberto",
    title = "{Sommerfeld enhancement: General results from field theory diagrams}",
    eprint = "0902.0688",
    archivePrefix = "arXiv",
    primaryClass = "hep-ph",
    doi = "10.1088/1126-6708/2009/05/024",
    journal = "JHEP",
    volume = "05",
    pages = "024",
    year = "2009"
}

@article{Cassel:2009wt,
    author = "Cassel, S.",
    title = "{Sommerfeld factor for arbitrary partial wave processes}",
    eprint = "0903.5307",
    archivePrefix = "arXiv",
    primaryClass = "hep-ph",
    reportNumber = "OUTP-0910P",
    doi = "10.1088/0954-3899/37/10/105009",
    journal = "J. Phys. G",
    volume = "37",
    pages = "105009",
    year = "2010"
}

@article{Slatyer:2009vg,
    author = "Slatyer, Tracy R.",
    title = "{The Sommerfeld enhancement for dark matter with an excited state}",
    eprint = "0910.5713",
    archivePrefix = "arXiv",
    primaryClass = "hep-ph",
    doi = "10.1088/1475-7516/2010/02/028",
    journal = "JCAP",
    volume = "02",
    pages = "028",
    year = "2010"
}

@article{Feng:2009hw,
    author = "Feng, Jonathan L. and Kaplinghat, Manoj and Yu, Hai-Bo",
    title = "{Halo Shape and Relic Density Exclusions of Sommerfeld-Enhanced Dark Matter Explanations of Cosmic Ray Excesses}",
    eprint = "0911.0422",
    archivePrefix = "arXiv",
    primaryClass = "hep-ph",
    reportNumber = "UCI-TR-2009-12",
    doi = "10.1103/PhysRevLett.104.151301",
    journal = "Phys. Rev. Lett.",
    volume = "104",
    pages = "151301",
    year = "2010"
}

@article{Arkani-Hamed:2008hhe,
    author = "Arkani-Hamed, Nima and Finkbeiner, Douglas P. and Slatyer, Tracy R. and Weiner, Neal",
    title = "{A Theory of Dark Matter}",
    eprint = "0810.0713",
    archivePrefix = "arXiv",
    primaryClass = "hep-ph",
    doi = "10.1103/PhysRevD.79.015014",
    journal = "Phys. Rev. D",
    volume = "79",
    pages = "015014",
    year = "2009"
}

@article{Hisano:2004ds,
    author = "Hisano, Junji and Matsumoto, Shigeki. and Nojiri, Mihoko M. and Saito, Osamu",
    title = "{Non-perturbative effect on dark matter annihilation and gamma ray signature from galactic center}",
    eprint = "hep-ph/0412403",
    archivePrefix = "arXiv",
    reportNumber = "ICRR-REPORT-513-2004-11, YITP-04-73",
    doi = "10.1103/PhysRevD.71.063528",
    journal = "Phys. Rev. D",
    volume = "71",
    pages = "063528",
    year = "2005"
}

@article{Coy:2022cpt,
    author = "Coy, Rupert and Xu, Xun-Jie and Yu, Bingrong",
    title = "{Neutrino forces and the Sommerfeld enhancement}",
    eprint = "2203.05455",
    archivePrefix = "arXiv",
    primaryClass = "hep-ph",
    doi = "10.1007/JHEP06(2022)093",
    journal = "JHEP",
    volume = "06",
    pages = "093",
    year = "2022"
}

@article{Choquette:2016xsw,
    author = "Choquette, Jeremie and Cline, James M. and Cornell, Jonathan M.",
    title = "{p-wave Annihilating Dark Matter from a Decaying Predecessor and the Galactic Center Excess}",
    eprint = "1604.01039",
    archivePrefix = "arXiv",
    primaryClass = "hep-ph",
    doi = "10.1103/PhysRevD.94.015018",
    journal = "Phys. Rev. D",
    volume = "94",
    number = "1",
    pages = "015018",
    year = "2016"
}

@article{Goodenough:2009gk,
    author = "Goodenough, Lisa and Hooper, Dan",
    title = "{Possible Evidence For Dark Matter Annihilation In The Inner Milky Way From The Fermi Gamma Ray Space Telescope}",
    eprint = "0910.2998",
    archivePrefix = "arXiv",
    primaryClass = "hep-ph",
    reportNumber = "FERMILAB-PUB-09-494-A",
    month = "10",
    year = "2009"
}

@article{Hooper:2010mq,
    author = "Hooper, Dan and Goodenough, Lisa",
    title = "{Dark Matter Annihilation in The Galactic Center As Seen by the Fermi Gamma Ray Space Telescope}",
    eprint = "1010.2752",
    archivePrefix = "arXiv",
    primaryClass = "hep-ph",
    reportNumber = "FERMILAB-PUB-10-414-A",
    doi = "10.1016/j.physletb.2011.02.029",
    journal = "Phys. Lett. B",
    volume = "697",
    pages = "412--428",
    year = "2011"
}

@article{Hooper:2011ti,
    author = "Hooper, Dan and Linden, Tim",
    title = "{On The Origin Of The Gamma Rays From The Galactic Center}",
    eprint = "1110.0006",
    archivePrefix = "arXiv",
    primaryClass = "astro-ph.HE",
    reportNumber = "FERMILAB-PUB-11-505-A",
    doi = "10.1103/PhysRevD.84.123005",
    journal = "Phys. Rev. D",
    volume = "84",
    pages = "123005",
    year = "2011"
}

@article{Abazajian:2012pn,
    author = "Abazajian, Kevork N. and Kaplinghat, Manoj",
    title = "{Detection of a Gamma-Ray Source in the Galactic Center Consistent with Extended Emission from Dark Matter Annihilation and Concentrated Astrophysical Emission}",
    eprint = "1207.6047",
    archivePrefix = "arXiv",
    primaryClass = "astro-ph.HE",
    doi = "10.1103/PhysRevD.86.083511",
    journal = "Phys. Rev. D",
    volume = "86",
    pages = "083511",
    year = "2012",
    note = "[Erratum: Phys.Rev.D 87, 129902 (2013)]"
}

@article{Zhou:2014lva,
    author = "Zhou, Bei and Liang, Yun-Feng and Huang, Xiaoyuan and Li, Xiang and Fan, Yi-Zhong and Feng, Lei and Chang, Jin",
    title = "{GeV excess in the Milky Way: The role of diffuse galactic gamma-ray emission templates}",
    eprint = "1406.6948",
    archivePrefix = "arXiv",
    primaryClass = "astro-ph.HE",
    doi = "10.1103/PhysRevD.91.123010",
    journal = "Phys. Rev. D",
    volume = "91",
    number = "12",
    pages = "123010",
    year = "2015"
}

@article{Fermi-LAT:2015sau,
    author = "Ajello, M. and others",
    collaboration = "Fermi-LAT",
    title = "{Fermi-LAT Observations of High-Energy $\gamma$-Ray Emission Toward the Galactic Center}",
    eprint = "1511.02938",
    archivePrefix = "arXiv",
    primaryClass = "astro-ph.HE",
    doi = "10.3847/0004-637X/819/1/44",
    journal = "Astrophys. J.",
    volume = "819",
    number = "1",
    pages = "44",
    year = "2016"
}

@article{Hisano:2002fk,
    author = "Hisano, Junji and Matsumoto, S. and Nojiri, Mihoko M.",
    title = "{Unitarity and higher order corrections in neutralino dark matter annihilation into two photons}",
    eprint = "hep-ph/0212022",
    archivePrefix = "arXiv",
    reportNumber = "ICRR-REPORT-495-2002-13, YITP-02-68",
    doi = "10.1103/PhysRevD.67.075014",
    journal = "Phys. Rev. D",
    volume = "67",
    pages = "075014",
    year = "2003"
}

@article{Hisano:2003ec,
    author = "Hisano, Junji and Matsumoto, Shigeki and Nojiri, Mihoko M.",
    title = "{Explosive dark matter annihilation}",
    eprint = "hep-ph/0307216",
    archivePrefix = "arXiv",
    reportNumber = "ICRR-REPORT-500-2003-4, YITP-03-42",
    doi = "10.1103/PhysRevLett.92.031303",
    journal = "Phys. Rev. Lett.",
    volume = "92",
    pages = "031303",
    year = "2004"
}

@article{Lattanzi:2008qa,
    author = "Lattanzi, Massimiliano and Silk, Joseph I.",
    title = "{Can the WIMP annihilation boost factor be boosted by the Sommerfeld enhancement?}",
    eprint = "0812.0360",
    archivePrefix = "arXiv",
    primaryClass = "astro-ph",
    doi = "10.1103/PhysRevD.79.083523",
    journal = "Phys. Rev. D",
    volume = "79",
    pages = "083523",
    year = "2009"
}

@article{Liu:2013vha,
    author = "Liu, Ze-Peng and Wu, Yue-Liang and Zhou, Yu-Feng",
    title = "{Sommerfeld enhancements with vector, scalar and pseudoscalar force-carriers}",
    eprint = "1305.5438",
    archivePrefix = "arXiv",
    primaryClass = "hep-ph",
    doi = "10.1103/PhysRevD.88.096008",
    journal = "Phys. Rev. D",
    volume = "88",
    pages = "096008",
    year = "2013"
}

@article{Blum:2016nrz,
    author = "Blum, Kfir and Sato, Ryosuke and Slatyer, Tracy R.",
    title = "{Self-consistent Calculation of the Sommerfeld Enhancement}",
    eprint = "1603.01383",
    archivePrefix = "arXiv",
    primaryClass = "hep-ph",
    doi = "10.1088/1475-7516/2016/06/021",
    journal = "JCAP",
    volume = "06",
    pages = "021",
    year = "2016"
}

@article{Ferrante:2025lbs,
    author = "Ferrante, Steven and Perelstein, Maxim and Yu, Bingrong",
    title = "{Sommerfeld Enhancement from Quantum Forces for Dark Matter}",
    eprint = "2507.12522",
    archivePrefix = "arXiv",
    primaryClass = "hep-ph",
    month = "7",
    year = "2025"
}

@article{Ittisamai:2025oxf,
    author = "Ittisamai, Pawin and Pongkitivanichkul, Chakrit and Sutwilai, Muhammaddaniya and Thongyoi, Nakorn and Uttayarat, Patipan",
    title = "{Effect of Cosmic Neutrino Background on the Dark Matter Self-interaction via Neutrino force}",
    eprint = "2509.20170",
    archivePrefix = "arXiv",
    primaryClass = "hep-ph",
    month = "9",
    year = "2025"
}

@article{Cheng:2025fak,
    author = "Cheng, Yu and Ge, Shuailiang",
    title = "{Background-Enhanced Axion Force by Axion Dark Matter}",
    eprint = "2504.02702",
    archivePrefix = "arXiv",
    primaryClass = "hep-ph",
    month = "4",
    year = "2025"
}

@article{Gan:2025icr,
    author = "Gan, Xucheng and Kim, Hyungjin and Mitridate, Andrea",
    title = "{Probing Quadratically Coupled Ultralight Dark Matter with Pulsar Timing Arrays}",
    eprint = "2510.13945",
    archivePrefix = "arXiv",
    primaryClass = "hep-ph",
    reportNumber = "DESY-25-134",
    month = "10",
    year = "2025"
}

@article{Kong:2025ccv,
    author = "Kong, Chuiyang and Di Mauro, Mattia",
    title = "{A Comprehensive Study of WIMP Models Explaining the Fermi-LAT Galactic Center Excess}",
    eprint = "2511.21808",
    archivePrefix = "arXiv",
    primaryClass = "hep-ph",
    month = "11",
    year = "2025"
}

@article{Cirelli:2008pk,
    author = "Cirelli, Marco and Kadastik, Mario and Raidal, Martti and Strumia, Alessandro",
    title = "{Model-independent implications of the e+-, anti-proton cosmic ray spectra on properties of Dark Matter}",
    eprint = "0809.2409",
    archivePrefix = "arXiv",
    primaryClass = "hep-ph",
    reportNumber = "IFUP-TH-2008-27, SACLAY-T08-139",
    doi = "10.1016/j.nuclphysb.2008.11.031",
    journal = "Nucl. Phys. B",
    volume = "813",
    pages = "1--21",
    year = "2009",
    note = "[Addendum: Nucl.Phys.B 873, 530--533 (2013)]"
}

@article{Pospelov:2008jd,
    author = "Pospelov, Maxim and Ritz, Adam",
    title = "{Astrophysical Signatures of Secluded Dark Matter}",
    eprint = "0810.1502",
    archivePrefix = "arXiv",
    primaryClass = "hep-ph",
    doi = "10.1016/j.physletb.2008.12.012",
    journal = "Phys. Lett. B",
    volume = "671",
    pages = "391--397",
    year = "2009"
}

@article{Fox:2008kb,
    author = "Fox, Patrick J. and Poppitz, Erich",
    title = "{Leptophilic Dark Matter}",
    eprint = "0811.0399",
    archivePrefix = "arXiv",
    primaryClass = "hep-ph",
    reportNumber = "FERMILAB-PUB-08-505-T",
    doi = "10.1103/PhysRevD.79.083528",
    journal = "Phys. Rev. D",
    volume = "79",
    pages = "083528",
    year = "2009"
}

@article{Pieri:2009zi,
    author = "Pieri, Lidia and Lattanzi, Massimiliano and Silk, Joseph",
    title = "{Constraining the Sommerfeld enhancement with Cherenkov telescope observations of dwarf galaxies}",
    eprint = "0902.4330",
    archivePrefix = "arXiv",
    primaryClass = "astro-ph.HE",
    doi = "10.1111/j.1365-2966.2009.15388.x",
    journal = "Mon. Not. Roy. Astron. Soc.",
    volume = "399",
    pages = "2033",
    year = "2009"
}

@article{Bovy:2009zs,
    author = "Bovy, Jo",
    title = "{Substructure Boosts to Dark Matter Annihilation from Sommerfeld Enhancement}",
    eprint = "0903.0413",
    archivePrefix = "arXiv",
    primaryClass = "astro-ph.HE",
    doi = "10.1103/PhysRevD.79.083539",
    journal = "Phys. Rev. D",
    volume = "79",
    pages = "083539",
    year = "2009"
}

@article{Yuan:2009bb,
    author = "Yuan, Qiang and Bi, Xiao-Jun and Liu, Jia and Yin, Peng-Fei and Zhang, Juan and Zhu, Shou-Hua",
    title = "{Clumpiness enhancement of charged cosmic rays from dark matter annihilation with Sommerfeld effect}",
    eprint = "0905.2736",
    archivePrefix = "arXiv",
    primaryClass = "astro-ph.HE",
    doi = "10.1088/1475-7516/2009/12/011",
    journal = "JCAP",
    volume = "12",
    pages = "011",
    year = "2009"
}

@article{Cholis:2010px,
    author = "Cholis, Ilias and Goodenough, Lisa",
    title = "{Consequences of a Dark Disk for the Fermi and PAMELA Signals in Theories with a Sommerfeld Enhancement}",
    eprint = "1006.2089",
    archivePrefix = "arXiv",
    primaryClass = "astro-ph.HE",
    doi = "10.1088/1475-7516/2010/09/010",
    journal = "JCAP",
    volume = "09",
    pages = "010",
    year = "2010"
}

@article{Zavala:2009mi,
    author = "Zavala, Jesus and Vogelsberger, Mark and White, Simon D. M.",
    title = "{Relic density and CMB constraints on dark matter annihilation with Sommerfeld enhancement}",
    eprint = "0910.5221",
    archivePrefix = "arXiv",
    primaryClass = "astro-ph.CO",
    doi = "10.1103/PhysRevD.81.083502",
    journal = "Phys. Rev. D",
    volume = "81",
    pages = "083502",
    year = "2010"
}

@article{Hryczuk:2011tq,
    author = "Hryczuk, Andrzej",
    title = "{The Sommerfeld enhancement for scalar particles and application to sfermion co-annihilation regions}",
    eprint = "1102.4295",
    archivePrefix = "arXiv",
    primaryClass = "hep-ph",
    doi = "10.1016/j.physletb.2011.04.016",
    journal = "Phys. Lett. B",
    volume = "699",
    pages = "271--275",
    year = "2011"
}

@article{Gan:2025nlu,
    author = "Gan, Xucheng and Liu, Da and Liu, Di and Luo, Xuheng and Yu, Bingrong",
    title = "{Detecting Ultralight Dark Matter with Matter Effect}",
    eprint = "2504.11522",
    archivePrefix = "arXiv",
    primaryClass = "hep-ph",
    reportNumber = "DESY-25-060",
    month = "4",
    year = "2025"
}

@article{McDonald:2012nc,
    author = "McDonald, Kristian L.",
    title = "{Sommerfeld Enhancement from Multiple Mediators}",
    eprint = "1203.6341",
    archivePrefix = "arXiv",
    primaryClass = "hep-ph",
    doi = "10.1007/JHEP07(2012)145",
    journal = "JHEP",
    volume = "07",
    pages = "145",
    year = "2012"
}

@article{Zhang:2013qza,
    author = "Zhang, Zhentao",
    title = "{Multi-Sommerfeld enhancement in dark sector}",
    eprint = "1307.2206",
    archivePrefix = "arXiv",
    primaryClass = "hep-ph",
    doi = "10.1016/j.physletb.2014.05.054",
    journal = "Phys. Lett. B",
    volume = "734",
    pages = "188--192",
    year = "2014",
    note = "[Erratum: Phys.Lett.B 774, 724--724 (2017)]"
}

\end{document}